\documentclass[11pt,a4paper]{article}
\usepackage[hyperref]{naaclhlt2018}
\usepackage{times}
\usepackage{enumitem}
\usepackage{latexsym}
\usepackage{multirow,pbox,booktabs}
\usepackage[export]{adjustbox}
\usepackage{url}
\usepackage[textsize=scriptsize]{todonotes}
\usepackage{flushend}
\usepackage{graphicx}

\makeatletter \let\c@table\c@figure \makeatother
\aclfinalcopy 
\title{CL Scholar: The ACL Anthology Knowledge Graph Miner}

\author{Mayank Singh, Pradeep Dogga\thanks{*These authors contributed equally to the study.}, Sohan Patro\footnotemark[1], Dhiraj Barnwal\footnotemark[1], Ritam Dutt\footnotemark[1],  \\
\bf{Rajarshi Haldar, Pawan Goyal and Animesh Mukherjee}\\
Department of Computer Science and Engineering\\
Indian Institute of Technology, Kharagpur, WB, India\\
mayank.singh@cse.iitkgp.ernet.in\\
}

\date{}

\begin{document}
\maketitle
\begin{abstract}
We present \textit{CL Scholar}, the ACL Anthology knowledge graph miner to facilitate high-quality search and exploration of current research progress in the computational linguistics community. In contrast to previous works, periodically crawling, indexing and processing of new incoming articles is completely automated in the current system. \textit{CL Scholar} utilizes both textual and network information for knowledge graph construction. As an additional novel initiative, \textit{CL Scholar} supports more than 1200 scholarly natural language queries along with standard keyword-based search on constructed knowledge graph. It answers \textit{binary}, \textit{statistical} and \textit{list} based natural language queries. The current system is deployed at \url{http://cnerg.iitkgp.ac.in/aclakg}. We also provide REST API support along with bulk download facility. Our code and data are available at \url{https://github.com/CLScholar}.
\end{abstract}

\section{Introduction}
ACL Anthology\footnote{https://aclweb.org/anthology/} is one of the popular initiatives of the Association for Computational  Linguistics (ACL) to curate all publications related to  computational linguistics and natural language processing at one common place. At present, it hosts more than 44,000 papers and is actively updated and maintained by Min Yen Kan. Since its inception, ACL Anthology functions as a repository with the collection of papers from ACL and related organizations in computational linguistics. However, it does not provide any additional statistics about authors, papers, venues, and topics. Also, it lacks advance search features such as article ranking by factoring in popularity or relevance, natural language query support, author profiles, topical search etc. 

\subsection{Previous systems built on ACL anthology}
Owing to above limitations, ACL anthology remained an archival repository for quite a long time. \citet{bird2008acl} developed the \textit{ACL Anthology Reference Corpus (ACL ARC)} as a collaborative attempt to provide a standardized testbed reference corpus based on the ACL Anthology. Later, \citet{radev2009acl} have invested humongous manual efforts to construct \textit{The {ACL} Anthology Network Corpus (AAN)}. AAN consists of a manually curated database of citations, collaborations, and summaries and statistics about the network. They have utilized two OCR processing tools PDFBox\footnote{https://pdfbox.apache.org/} and ParsCit~\cite{councill2008parscit} for curation. AAN was continuously updated till  2013~\cite{radev2013}. Recently, this project has been moved to Yale University as part of the new LILY group\footnote{http://tangra.cs.yale.edu/newaan/}.

\subsection{The computational linguistic knowledge graph}
As a similar initiative, in this paper, we demonstrate the development of \textit{CL Scholar} which automatically mines ACL anthology and constructs computational linguistic knowledge graph (hereafter \textit{`CLKG'}). The current framework automatically crawls new articles, processes, indexes, constructs knowledge graph and generates searchable statistics without involving tedious manual annotations. We leverage state-of-the-art scientific article processing tool OCR++~\cite{singh-EtAl:2016:COLING2} for robust and automatic information extraction from scientific articles. OCR++ is an open-source framework that can extract from scholarly articles the metadata, the structure and the bibliography.

\begin{figure*}[!tbh]
\centering
\includegraphics[width=0.9\hsize]{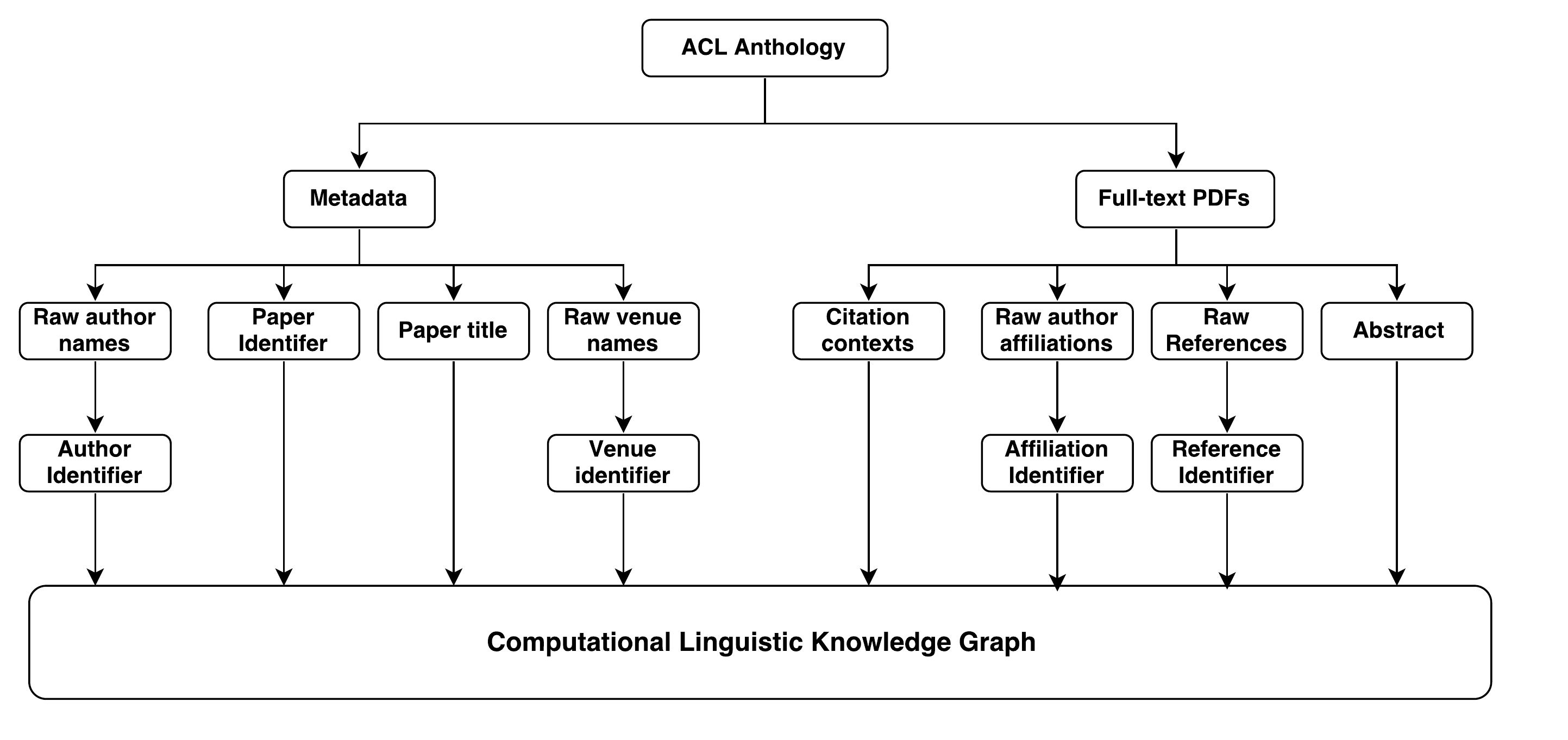}
\vspace{-0.7cm}
\caption{Data processing flow diagram.}
\label{fig:knowledge_graph}
\vspace{-0.5cm}
\end{figure*}

The constructed $CLKG$ is modeled as a heterogeneous graph~\cite{sun2009ranking} consisting of four entities: author, paper, venue, and field. We utilize metapaths~\cite{sun2012mining} to implement the query retrieval framework. 

\subsection{Natural language queries}

In the first-of-its-kind initiative, we extend the functionalities of \textit{CL Scholar} to answer natural language queries (hereafter `\textit{NLQ}') along with standard keyword-based queries. Currently, it answers \textit{binary}, \textit{statistical} and \textit{list} based $NLQ$. Overall, we handle more than 1200 variations of $NLQ$.  

\noindent\textbf{Outline:} The rest of the paper is organized as follows. Section~\ref{sec:dataset} describes the ACL Anthology dataset. Section~\ref{sec:preprocess} details step by step extraction procedure for $CLKG$ construction. In section \ref{sec:knowledge_graph}, we describe $CLKG$. We describe our framework in section~\ref{sec:framework}. We conclude in section~\ref{sec:end} and identify future work.

 \vspace{-0.2cm}
\begin{table}[!tbh]
\centering
  \caption{General statistics about the ACL Anthology dataset.}
  \label{tab:dataset}
  \begin{tabular}{lc} \toprule
  Number of papers &42,069\\ 
  Year range& 1965--2017 \\ 
  Total authors& 37,752\\
  Total unique authors& 33,372\\
  Total unified venues& 33\\
  \bottomrule \hline
 \end{tabular}
 \vspace{-0.2cm}
\end{table}

\section{Dataset}
\label{sec:dataset}
\textit{CL Scholar} uses metadata and full-text PDF research articles crawled from ACL Anthology. ACL Anthology consists of more than 40,000 research articles published in more than 33 computational linguistic events (venues) including conferences, workshops, and journals. Table~\ref{tab:dataset}  presents general statistics of the crawled dataset.

We crawl both metadata information (unique article identifier, article title, authors' names, and venue) as well as full-text PDF articles. Next, we describe in detail several pre-processing steps and knowledge graph construction methodology.

\section{Pre-processing and knowledge graph construction}
\label{sec:preprocess}
We process full-text PDFs using state-of-the-art extraction tool \textit{OCR++}~\cite{singh-EtAl:2016:COLING2}. We extract references, citation contexts, author affiliations and URLs from full-text. \textit{OCR++} also provides reference to citation contexts mapping. Raw information with several variations like author names, venue names and affiliations are assigned unique identifiers using standard indexing approaches. We only consider those reference papers that are present in ACL anthology. This rich textual, as well as citation relationship information, is utilized in the construction of $CLKG$. Figure~\ref{fig:knowledge_graph} presents the $CLKG$ construction from metadata and full-text PDF files crawled from ACL anthology. 

\section{Computational linguistic knowledge graph}
\label{sec:knowledge_graph}
Computational linguistic knowledge graph ($CLKG$) is a \textit{heterogeneous graph}~\cite{sun2009ranking} consisting of four entities: author ($A$), paper ($P$), venue ($V$) and field ($F$) as nodes. Each entity is associated with few properties, for example, properties of $P$ are publication year, title, abstract, etc. Similarly, properties of $A$ are name, publication trend, affiliation etc. We utilize \textit{metapaths}~\cite{sun2012mining} between  entities to express semantic relations. For example, simple metapaths like $A$$\rightarrow$$P$ and $V$$\rightarrow$$P$ represent ``author of'' and ``published at'' relations respectively, whereas complex metapaths like  $V$$\rightarrow$$A$$\rightarrow$$P$ and $F$$\rightarrow$$A$$\rightarrow$$P$ represent ``authors of papers published at'' and ``authors of papers in'' relations respectively. We leverage metapaths to develop \textit{CL Scholar} (described in the next section).

\vspace{-0.5cm}
\begin{figure}[!tbh]
\centering
\includegraphics[trim=0 0 0 15,clip,width=0.9\hsize]{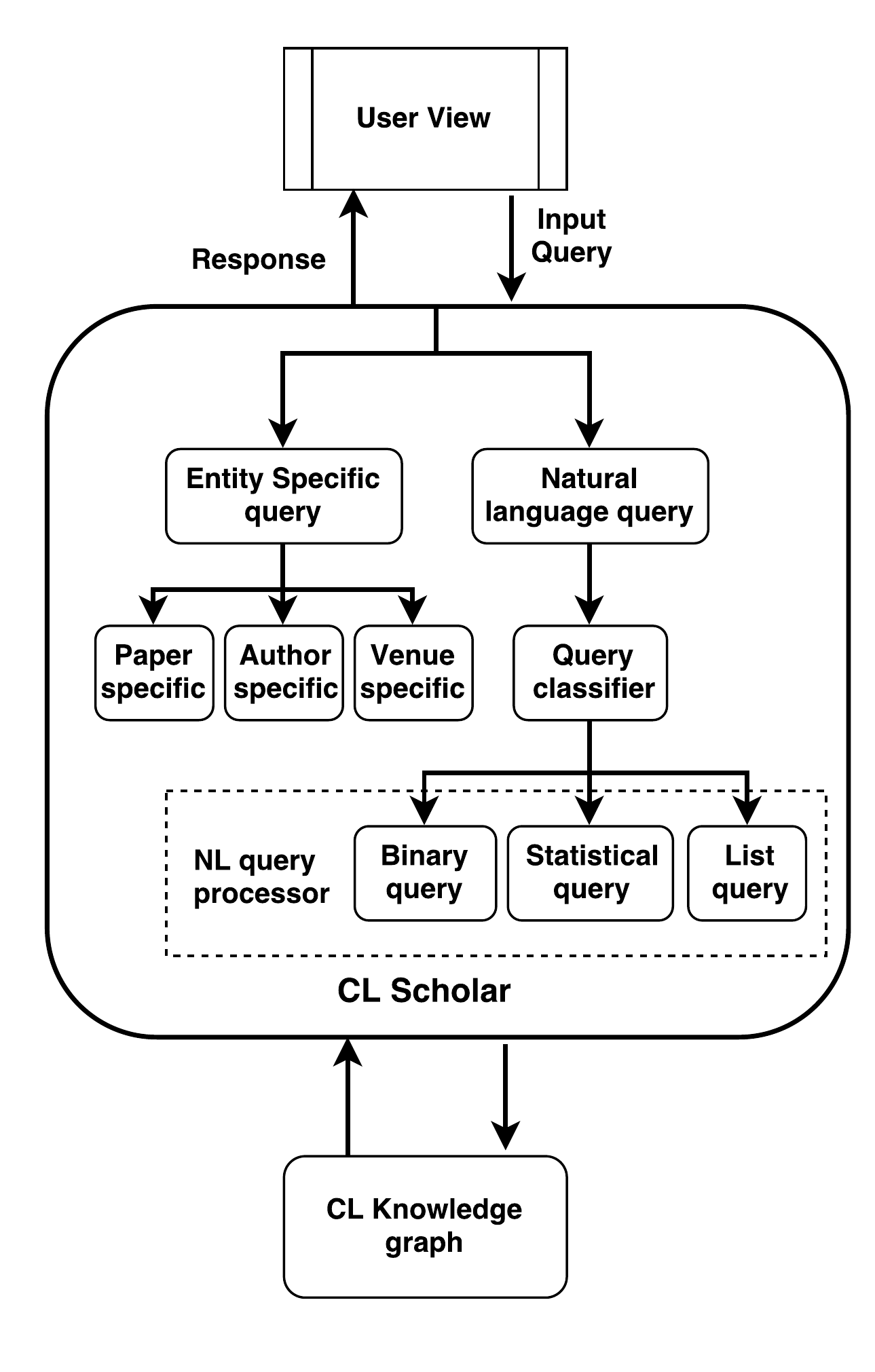}
\vspace{-1cm}
\caption{CL Scholar framework.}
\label{fig:framwork}
\vspace{-0.5cm}
\end{figure}

\section{CL Scholar}
\label{sec:framework}
\textit{CL scholar} fetches information from $CLKG$ as per the input query from the user. The current framework is divided into two modules -- 1) natural language based query retrieval, and 2) entity specific query retrieval. Figure~\ref{fig:framwork} shows \textit{CL Scholar} framework.

\begin{table*}[!tbh]
\centering
\caption{Representative queries from three natural language query classes. $A$ represents author, $P$ represents paper, $V$ represents venue and $F$ represents field. The list of supported queries is available online at \textit{CL Scholar} portal.} \label{tab:NL_query}
\resizebox{0.95\textwidth}{!}{\begin{minipage}{\textwidth}
\begin{tabular}{clll} \toprule
&\multicolumn{1}{c}{\textbf{Binary queries}} & \multicolumn{1}{c}{\textbf{Statistical queries}} &\multicolumn{1}{c}{ \textbf{List queries}}\\
\midrule
1.&Is $V$ accepting papers from $F$&
\parbox[t]{4.5cm}{How many $F$ papers are published in $V$ over the years}&\parbox[t]{4.5cm}{List the papers from $F$ accepted in $V$}\\
2.&Has $A$ published any paper&
\parbox[t]{4.5cm}{How many papers are published by $A$}&\parbox[t]{4.5cm}{List the papers published by $A$}\\
3.&Does $A$ publish papers on $F$&
\parbox[t]{4.5cm}{How many papers are published by $A$ in $F$}&\parbox[t]{4.5cm}{List the papers published by $A$ on $F$}\\
4.&\parbox[t]{4.5cm}{Are there any papers published by $A_1$ and $A_2$ together}&
\parbox[t]{4.5cm}{How many papers are published by $A_1$ and $A_2$ together}&\parbox[t]{4.5cm}{List the papers published by $A1$ and $A2$ together}\\
5.& \parbox[t]{4.5cm}{Are there any papers of $A$ with positive sentiment}&
\parbox[t]{4.5cm}{How many papers are there of $A$ with positive sentiment}&\parbox[t]{4.5cm}{List of papers with positive sentiment of  $A$}\\
\bottomrule
\hline
\end{tabular}
\end{minipage}}
\vspace{-0.4cm}
\end{table*}

\subsection{Natural language query retrieval}
\label{sec:nlp_query}
The first module answers natural language queries ($NLQ$). It consists of two sub-modules, 1) the query classifier, and 2) the NL query processor.  \textit{Query classifier} classifies user input into one of the three basic types of $NLQ$ using regular expression patterns. \textit{NL query processor} processes query based on its type determined by query classifier. Given an input natural language query, we utilize longest subsequence match to identify entity instances. The three types of $NLQ$ are:
\begin{enumerate}[noitemsep,nolistsep]
\item \textbf{Binary queries:} These represent a set of queries for which user demands a `yes' or `no' type answer. Table~\ref{tab:NL_query} lists few interesting binary queries. 
\item \textbf{Statistical queries:}  These represent set of queries which the knowledge base returns with some statistics. Currently, we support three types of statistics -- 1) temporal, 2) cumulative, and 3) comparison. Temporal represents year-wise statistics, cumulative represents overall statistics and comparison represents comparative statistics between two or more instances of the same entity type. Table~\ref{tab:NL_query} lists few representative statistical queries.  
\item \textbf{List queries:} These represent set of queries for which the knowledge base returns a list of papers, authors or venues.  Table~\ref{tab:NL_query} also enumerates few representative list queries.  
\end{enumerate}

\subsection{Entity specific query retrieval}
\textit{CL scholar} also supports entity specific retrieval. As described in section~\ref{sec:knowledge_graph}, $CLKG$ consists of four entities: paper, author, venue, and field. Currently, our system supports three\footnote{The fourth sub-module is still under development.} entity specific retrieval schemes handled by three sub-modules: 

\begin{enumerate}[noitemsep,nolistsep]
\item \textbf{Paper specific:} This sub-module returns paper specific information. Currently, we retrieve and display author names and affiliations, abstract, publication year and venue, cumulative and year-wise citations, list of references, citer papers, co-cited papers present in ACL anthology and list of URLs present in the paper text. We also show average sentiment score received by the queried paper by utilizing incoming citation contexts. Table~\ref{tab:entity_specific_query} shows three representative paper specific queries. 
\item \textbf{Author specific:} This sub-module handles author specific queries. Given an author name, the system shows its cumulative and year-wise publication and citation count, collaborator list with an average number of collaborations, current and temporal H-index and temporal topic distribution. We also list author's publications in ACL anthology. Table~\ref{tab:entity_specific_query} lists three author specific queries with first name, last name and full name respectively.
\item \textbf{Venue specific:} We also answer venue specific queries. For each venue specific query, the system shows cumulative and year-wise publication and citation count, 2-year impact factor, recently held year and list of collaborating venues. Table~\ref{tab:entity_specific_query} shows three representative venue specific queries. 
\end{enumerate}

\vspace{-0.3cm}
\begin{table}[!tbh]
\centering
\caption{Representative entity specific queries.} \label{tab:entity_specific_query}
\begin{tabular}{ccc} \toprule
\parbox[t]{1.5cm}{\centering \textbf{Paper specific}} & \parbox[t]{1.5cm}{\centering \textbf{Author specific}}&\parbox[t]{1.5cm}{\centering \textbf{Venue specific}}\\
\midrule
OCR&Chris&NAACL\\
Deep learning&Singh&SIGDAT\\
Word embeddings&Aravind Joshi&ACL
\\
\bottomrule
\hline
\end{tabular}
\vspace{-0.3cm}
\end{table}

\subsection{Additional insights}
\label{sec:insights}
We provide two additional insights by analyzing incoming citation contexts. First, we present a summary generated from incoming the citation contexts~\cite{Qazvinian:2008:SPS:1599081.1599168}. Currently, we show five summary sentences against each paper. Second, we also compute sentiment score of each citation context by leveraging a standard sentiment analyzer~\cite{Athar:2012:CCS:2382029.2382125}. We aggregate by averaging over the sentiment score of all the incoming citation contexts. 

\subsection{Ranking}
Currently, we employ popularity based ranking of retrieved results. We utilize current citation count as a measure of popularity. In future, we plan to deploy other ranking schemes like recency, impact, sentiment, relevance, etc.  

\subsection{Deployment}
\begin{figure}[!tbh]
\centering
\includegraphics[width=1\hsize]{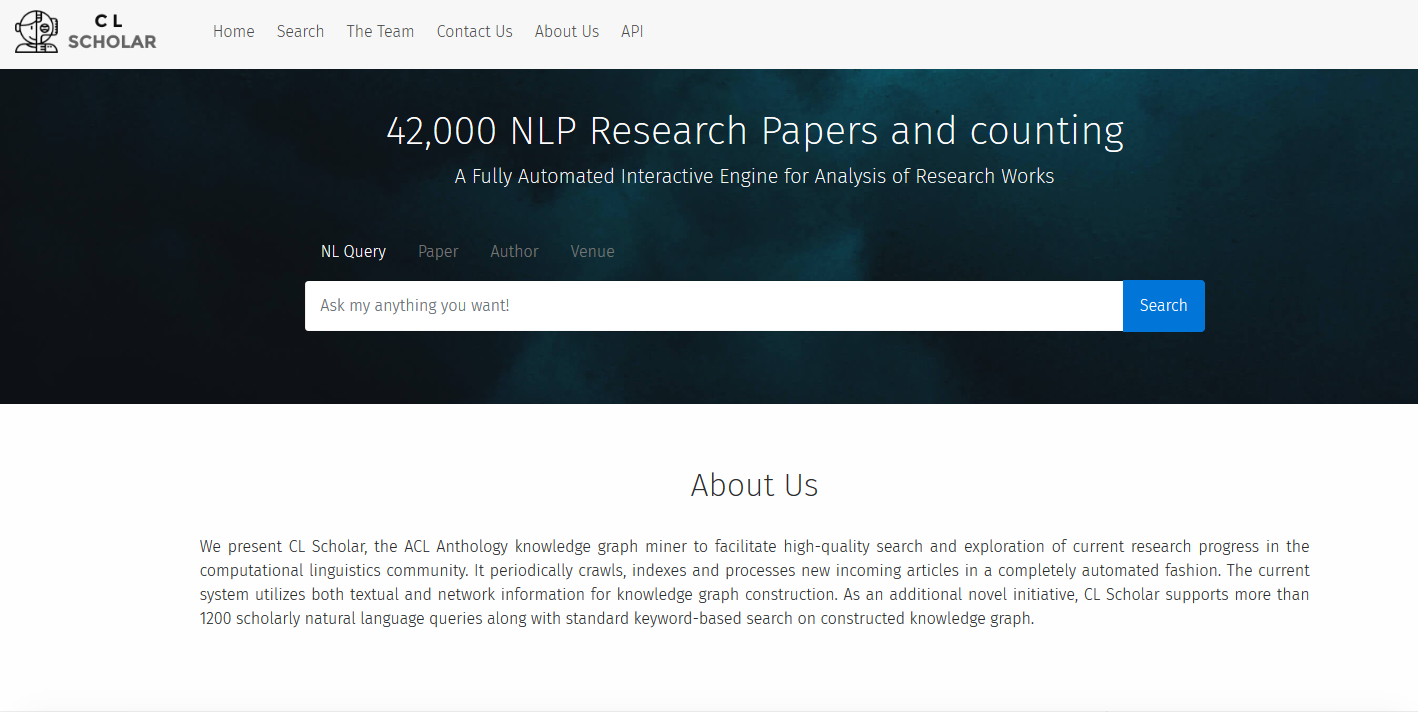}
\vspace{-0.7cm}
\caption{Snapshot of CL Scholar landing page.}
\label{fig:portal}
\vspace{-0.3cm}
\end{figure}

\textit{CL Scholar} is developed using ReactJS framework. The system also supports REST API requests which are powered by a NodeJS server with data being served using MongoDB. It is currently accessible at our research group page\footnote{http://cnerg.iitkgp.ac.in/aclakg}. More information about API usage is available at API support page\footnote{http://cnerg.iitkgp.ac.in/aclakg/api}. In addition, the entire knowledge graph can also be easily downloaded in a plain text format. Figure~\ref{fig:portal} shows a snapshot of the \textit{CL Scholar} landing page.

The current system is still under development. Currently, we assume that spellings are correct for NLQ. We do not support instant query search. We also do not support query recommendations.

\section{Conclusion}
\label{sec:end}
In this paper, we propose a fully automatic approach for the development of computational linguistic knowledge graph from full-text PDF articles available in ACL Anthology.  We also develop first-of-its-kind academic natural language query retrieval system. Currently, our system can answer three different types of natural language queries. In future, we plan to extend the query set. We also plan to append structural information within knowledge graphs such as section labeling of citations, figure and table captions etc. We also plan to conduct extensive evaluation to compare \textit{CL Scholar} with state-of-the-art systems. 

\newpage
\bibliographystyle{acl_natbib}
\bibliography{naacl}
\flushend

\end{document}